\documentclass[12pt]{iopart} 

\begin{document}

\title{The present universe 
in the Einstein frame, metric-affine $R+1/R$ gravity} 

\author{Nikodem J Pop\l awski}

\address{Department of Physics, Indiana University, 
727 East Third Street, Bloomington, IN 47405, USA}
\ead{nipoplaw@indiana.edu}

\begin{abstract}
We study the present, flat isotropic universe in $1/R$-modified gravity.
We use the Palatini (metric-affine) variational principle and
the Einstein (metric-compatible connected) conformal frame.
We show that the energy density scaling deviates from the usual scaling
for nonrelativistic matter, and the largest deviation
occurs in the present epoch.
We find that the current deceleration parameter derived from
the apparent matter density parameter is consistent with observations.
There is also a small overlap between the predicted and observed values
for the redshift derivative of the deceleration parameter.
The predicted redshift of the deceleration-to-acceleration transition
agrees with that in the $\Lambda CDM$ model but it is larger than the value
estimated from SNIa observations.

\end{abstract}


\maketitle

\section{Introduction} 
The most accepted explanation of current cosmic 
acceleration~\cite{SN,CMB1,CMB2} is that the 
universe is dominated by dark energy~\cite{DE}.
However, it is also possible to modify Einstein's general relativity 
to obtain gravitational field equations that allow accelerated
expansion. 
A particular class of alternative theories of gravity that has recently 
attracted a lot of interest is that of the $f(R)$ gravity 
models, in which the gravitational Lagrangian is a function of 
the curvature scalar $R$~\cite{Bar}. 
It has been shown that cosmic acceleration may originate from the addition of 
a term $R^{-1}$ (or other negative powers of $R$) to the Einstein-Hilbert
Lagrangian $R$~\cite{fR1,fR2,Pal2,Pal8,Pal5,Pal6}.

As in general relativity, these models obtain the field equations by 
varying the total action for both the field and matter. 
In this paper, we use the metric-affine (Palatini) variational principle,
according to which the metric and connection are considered as 
geometrically independent quantities, and the action is varied 
with respect to both of them~\cite{Pal1}. 
The other one is the metric (Einstein-Hilbert) variational principle, 
according to which the action is varied with respect to the metric 
tensor $g_{\mu\nu}$, and the affine connection coefficients are
the Christoffel symbols of $g_{\mu\nu}$. 
Both approaches give the same result 
only if we use the standard Einstein-Hilbert action.
The field equations in the Palatini formalism are second-order 
differential equations, while for metric theories they are 
fourth-order~\cite{eq1}. 
Another remarkable property of the metric-affine approach is that the 
field equations in vacuum reduce to the standard Einstein equations of 
general relativity with a cosmological constant~\cite{univ}.

One can show that any of these theories of gravitation 
is conformally equivalent to the Einstein theory
of the gravitational field interacting with additional matter 
fields~\cite{eq2}. 
This can be done by means of a Legendre 
transformation, which in classical mechanics replaces 
the Lagrangian of a mechanical system with the Helmholtz 
Lagrangian~\cite{eq3}. 
For an $f(R)$ gravity, the scalar degree of freedom 
due to the occurrence of nonlinear second-order terms in the Lagrangian
is transformed into an auxiliary scalar field $\phi$. 
The set of variables $(g_{\mu\nu},\,\phi)$ is commonly called the 
{\it Jordan conformal frame}. 

In the Jordan frame, the connection is not metric compatible.
The compatibility can be restored by a certain conformal 
transformation of the metric: $g_{\mu\nu}\rightarrow 
h_{\mu\nu}=f'(R)g_{\mu\nu}$. 
The new set $(h_{\mu\nu},\,\phi)$ is called the {\it Einstein conformal 
frame}. 
Although both frames are equivalent mathematically, they are {\it not}  
equivalent physically~\cite{EJ2}, and 
the interpretation of cosmological observations can drastically 
change depending on the adopted frame~\cite{EJ1}.
In the Palatini formalism, which frame is physical is a matter of choice,
although this question should be ultimately answered by experiment 
or observation.

The consistence of metric-affine $f(R)$ gravities
with cosmological observations has already been studied for the
Jordan frame~\cite{obs}.
In this paper, we regard the Einstein conformal frame as the physical one.
We assume that the gravitational Lagrangian contains
the usual Einstein-Hilbert part $R$ and the $1/R$ term which causes
current cosmic acceleration~\cite{fR2}.
In section~2, we review the metric-affine formalism for an $f(R)$ 
gravity in the Einstein frame~\cite{Niko}. 
In section~3, we apply the gravitational field equations for the $R+1/R$ 
Lagrangian to a homogeneous and isotropic universe, and solve them
for the early and late universe in the quadratic approximation.
In section~4, we derive the relations between the present deceleration
parameter $q_0$, the matter density parameter $\Omega_M$ and the redshift
of the deceleration-to-acceleration transition $z_t$ in the Einstein
frame of the metric-affine $1/R$-modified gravity.
We also compare these relations with cosmological observations.
The results are summarized in section~5.

\section{Metric-affine formalism in the Einstein conformal frame}
The action for an $f(R)$ gravity in the original (Jordan) frame 
with the metric $\tilde{g}_{\mu\nu}$ is given by
\begin{equation}
S_J=-\frac{1}{2\kappa c}\int d^4 
x\bigl[\sqrt{-\tilde{g}}L(\tilde{R})\bigr] + 
S_m(\tilde{g}_{\mu\nu},\psi).
\label{action1}
\end{equation}
Here, $\sqrt{-\tilde{g}}L(\tilde{R})$ is a Lagrangian density that depends 
on the curvature scalar 
$\tilde{R}=R_{\mu\nu}(\Gamma^{\,\,\lambda}_{\rho\,\sigma})\tilde{g}^{\mu\nu}$, 
$S_m$ is the action for matter represented 
symbolically by $\psi$ and independent of the connection, 
and $\kappa=\frac{8\pi G}{c^4}$. 
Tildes indicate quantities calculated in the Jordan frame.

Variation of the action $S_J$ with respect to $\tilde{g}_{\mu\nu}$ 
yields the field equations
\begin{equation}
L'(\tilde{R})R_{\mu\nu}-\frac{1}{2}L(\tilde{R})\tilde{g}_{\mu\nu}=\kappa 
T_{\mu\nu},
\label{field1}
\end{equation} 
where the dynamical energy-momentum tensor of matter is generated 
by the Jordan metric tensor:
\begin{equation}
\delta S_m=\frac{1}{2c}\int d^4 x\sqrt{-\tilde{g}}\,T_{\mu\nu}\delta\tilde{g}^{\mu\nu},
\label{EMT1}
\end{equation} 
and the prime denotes the derivative of a function with respect to its variable.
From variation of $S_J$ with 
respect to the connection $\Gamma^{\,\,\rho}_{\mu\,\nu}$,
it follows that this connection is given by the Christoffel 
symbols of the conformally transformed metric~\cite{Niko}: 
\begin{equation}
g_{\mu\nu}=L'(\tilde{R})\tilde{g}_{\mu\nu}.
\label{conf}
\end{equation}
The metric $g_{\mu\nu}$ defines the Einstein frame, in which the connection is metric compatible.

The action~(\ref{action1}) is dynamically equivalent 
to the following Helmholtz action~\cite{eq1,eq3}:
\begin{equation}
S_H=-\frac{1}{2\kappa c}\int d^4 x\sqrt{-\tilde{g}}\bigl[L(\phi(p))+p(\tilde{R}-\phi(p))\bigr]+S_m(\tilde{g}_{\mu\nu},\psi),
\label{action2}
\end{equation}
where $p$ is a new scalar variable.
The function $\phi(p)$ is determined by
\begin{equation}
\frac{\partial 
L(\tilde{R})}{\partial\tilde{R}}\bigg{\vert}_{\tilde{R}=\phi(p)}=p.
\label{phi}
\end{equation}
From equations~(\ref{conf}) and~(\ref{phi}), it follows that
\begin{equation}
\phi=RL'(\phi).
\label{resc}
\end{equation}

In the Einstein frame, the action~(\ref{action2}) becomes the standard 
Einstein-Hilbert action of general relativity with additional
scalar field:
\begin{equation}
S_E=-\frac{1}{2\kappa c}\int d^4 x\sqrt{-g}\bigl[R-\frac{\phi(p)}{p}+\frac{L(\phi(p))}{p^2}\bigr]+S_m(p^{-1}g_{\mu\nu},\psi).
\label{action3}
\end{equation}
Here, $R$ is the curvature scalar of the metric $g_{\mu\nu}$.
Choosing $\phi$ as the scalar variable leads to~\cite{Niko}
\begin{equation}
S_E=-\frac{1}{2\kappa c}\int d^4 x\sqrt{-g}\bigl[R-2V(\phi)\bigr]+S_m([L'(\phi)]^{-1}g_{\mu\nu},\psi),
\label{action4}
\end{equation}
where $V(\phi)$ is the effective potential,
\begin{equation}
V(\phi)=\frac{\phi L'(\phi)-L(\phi)}{2[L'(\phi)]^2}.
\label{pot}
\end{equation}

Variation of the action~(\ref{action4}) with respect to 
$g_{\mu\nu}$ yields the equation of the gravitational field in
the Einstein frame~\cite{Niko,Pal3}: 
\begin{equation}
R_{\mu\nu}-\frac{1}{2}Rg_{\mu\nu}=\frac{\kappa 
T_{\mu\nu}}{L'(\phi)}-V(\phi)g_{\mu\nu},
\label{EOF1}
\end{equation}
while variation with respect to $\phi$ reproduces equation~(\ref{resc}).
Contracting~(\ref{EOF1}) with the metric tensor $g^{\mu\nu}$ gives
\begin{equation}
R=-\frac{\kappa T}{L'(\phi)}+4V(\phi),
\label{struc1}
\end{equation} 
which is an equation for $R$ since both $\phi$ and $T=T_{\mu\nu}g^{\mu\nu}$ 
depend only on $R$ due to~(\ref{resc}) and~(\ref{struc1}). 
This is equivalent to~\cite{Niko}
\begin{equation}
\phi L'(\phi)-2L(\phi)=\kappa TL'(\phi).
\label{struc2}
\end{equation}

Let us consider the CDTT Lagrangian~\cite{fR2,Pal2} 
\begin{equation}
L(\phi)=\phi-\frac{\alpha^2}{3\phi},
\label{acc1}
\end{equation}
where $\alpha$ is a positive constant.\footnote[1]{
Equation~(\ref{phi}) states that $\phi$ is the curvature scalar in the Jordan
frame $\tilde{R}$. The Lagrangian~(\ref{acc1}) can thus be written as
$L(\tilde{R})=\tilde{R}-\frac{\alpha^2}{3\tilde{R}}$ and such a model is
referred to as the $R+1/R$ gravity or $1/R$-modified gravity.}\footnote[2]{
The Lagrangian of~\cite{Niko} has an additional term $R^3$, which
is the simplest way of introducing inflation in a metric-affine $f(R)$
gravity~\cite{R3}.
In this work, we omit this term since we are interested
in later epochs.}
Equation~(\ref{struc2}) for $T=0$ yields one de Sitter solution,
\begin{equation}
\phi_{ca}=-\alpha,
\label{acc2}
\end{equation}
describing present cosmic acceleration~\cite{Pal2}.
The corresponding value of the curvature scalar in the Einstein frame 
is determined by equation~(\ref{resc}):
\begin{equation}
R=\frac{\phi}{1+\frac{\alpha^2}{3\phi^2}},
\label{resc2}
\end{equation}
from which we obtain
\begin{equation}
R_{ca}=-\frac{3\alpha}{4}.
\label{acc3}
\end{equation}
Finally, the Einstein equations~(\ref{EOF1}) become
\begin{equation}
R_{\mu\nu}-\frac{1}{2}Rg_{\mu\nu}=\frac{\kappa 
T_{\mu\nu}}{1+\frac{\alpha^2}{3\phi^2}}-\frac{\frac{\alpha^2}{3\phi}}{(1+\frac{\alpha^2}{3\phi^2})^2}g_{\mu\nu}.
\label{field2}
\end{equation}
Equations~(\ref{resc2}) and~(\ref{field2}) give the relation between matter
and geometry in the Einstein frame of $1/R$-modified gravity.\footnote[3]{
The value of $\alpha$ is on the order of $10^{-53}\mbox{m}^{-2}$~\cite{Pal2}.}

\section{The field equations in the early and late universe}
We begin with the Einstein equations for the early
universe in which the $1/R$ term does not dominate.
In such a universe $\kappa T\gg\alpha$ and we can study the field
equations in the second
approximation of a small quantity $\frac{\alpha}{\kappa T}$.  
Equation~(\ref{struc2}) becomes cubic in $\phi$:
\begin{equation}
\phi^3+\kappa T\phi^2-\alpha^2\phi+\frac{\alpha^2\kappa T}{3}=0,
\label{struc3}
\end{equation}
and its solution which deviates from $-\kappa T$ (the solution for $\alpha=0$) 
by terms linear and quadratic in $\frac{\alpha}{\kappa T}$ is
\begin{equation}
\phi=-\kappa T-\frac{4\alpha^2}{3\kappa T}+O(\alpha^3).
\label{struc4}
\end{equation}
In this approximation, the Einstein equations become
\begin{equation}
R_{\mu\nu}-\frac{1}{2}Rg_{\mu\nu}=\kappa_{eff}T_{\mu\nu}+\Lambda_{eff}g_{\mu\nu},
\label{EOF2}
\end{equation}
where
\begin{equation}
\kappa_{eff}=\kappa\Bigl(1-\frac{\alpha^2}{3\kappa^2 T^2}\Bigr)
\label{effkappa}
\end{equation}
is the effective gravitational coupling constant and
\begin{equation}
\Lambda_{eff}=\frac{\alpha^2}{3\kappa T} 
\label{efflambda}
\end{equation}
is the effective cosmological constant.

We now proceed to the field equations in the early 
Friedmann-Lema\^{i}tre-Robertson-Walker (FLRW) universe.
Let us consider a homogeneous and isotropic universe which is spatially 
flat~\cite{CMB2}.
In this case, the interval is given by
\begin{equation}
ds^2=c^2dt^2-a^2(t)(dx^2+dy^2+dz^2),
\label{FLRW}
\end{equation}
where $a(t)$ is the scale factor.
Moreover, the energy-momentum tensor of matter in the comoving 
frame of reference is diagonal:
\begin{equation}
T_{\mu}^{\nu}=diag(\epsilon,\,-P,\,-P,\,-P),
\label{EMT2}
\end{equation} 
where $\epsilon$ is the energy density and $P$ denotes pressure.
For this metric and for the case of dust ($P=0$), equation~(\ref{EOF2}) has
two independent components:
\begin{eqnarray}
& & \frac{3\dot{a}^2}{c^2 a^2}=\kappa\Im, \\
& & \frac{\dot{a}^2+2a\ddot{a}}{c^2 a^2}=\kappa\wp,
\end{eqnarray}
where the generalized energy density $\Im$ in this approximation is equal
to the energy density $\epsilon$ and the generalized pressure is given by
\begin{equation}
\wp=-\frac{\alpha^2}{3\kappa^2\epsilon}.
\end{equation}
The dot denotes the differentiation with respect to time.

From the conservation law for the generalized quantities:
\begin{equation}
\dot{\Im}+3H(\Im+\wp)=0,
\label{cons}
\end{equation}
where $H=\frac{\dot{a}}{a}$ is the Hubble parameter,
we obtain the scaling law for the energy density:
\begin{equation}
\epsilon^2=\frac{E_0^2}{a^6}-\frac{\alpha^2}{3\kappa^2},
\label{scalen1}
\end{equation}
where $E_0$ is a positive constant proportional to the total energy
of matter within a sphere of radius $a$ in the limit 
$\frac{\alpha}{\kappa\epsilon}\rightarrow0$.
In the course of time, $a$ increases but the right-hand side
of~(\ref{scalen1}) never becomes negative since 
the assumption $\kappa T\gg\alpha$ ceases to hold when 
$a^3\sim\frac{E_0\kappa}{\alpha}$.
For $b(t)=a^{3/2}(t)$, we obtain
\begin{equation}
\frac{16}{3c^4}(\dot{b})^4=3\kappa^2 E_0^2-\alpha^2 b^4.
\label{scal}
\end{equation}

For a very late universe $\kappa T\ll\alpha$, and the Einstein equations
in the second approximation of a small quantity $\frac{\kappa T}{\alpha}$
are~\cite{Niko}
\begin{equation}
R_{\mu\nu}-\frac{1}{2}Rg_{\mu\nu}=\frac{3}{4}\kappa T_{\mu\nu}\Bigl(1+\frac{\kappa T}{3\alpha}\Bigr)+\Lambda g_{\mu\nu},
\label{EOF3}
\end{equation}
where $\Lambda=\frac{3\alpha}{16}-\frac{\kappa^2 T^2}{16\alpha}$.
The coupling between matter and the gravitational field 
is decreased by a factor which tends to $3/4$ as $T\rightarrow0$.
Equation~(\ref{EOF3}) can be rewritten as
\begin{equation}
\frac{R_{\mu\nu}-g_{\mu\nu}(\frac{R}{3}-\frac{R^2}{9\alpha}+\frac{\alpha}{8})}{\frac{1}{2}-\frac{R}{3\alpha}}=\kappa T_{\mu\nu}.
\label{EOF4}
\end{equation}
From the same equation it also follows that the energy-momentum tensor 
in the Einstein frame is not covariantly conserved:~\footnote[4]{
It might seem that this tensor is not conserved for $\alpha=0$.
In this case, however, $T=0$ since $\kappa T\ll\alpha$.}
\begin{equation}
T_{\mu\nu}^{\,\,\,\,\,;\nu}=\frac{\frac{\kappa}{4\alpha}(T_{\mu\nu}T^{;\nu}-\frac{1}{2}TT_{;\mu})}{\frac{3}{4}+\frac{\kappa T}{4\alpha}},
\label{EOF5}
\end{equation}
unless $T_{\mu\nu}-\frac{1}{2}Tg_{\mu\nu}=0$.\footnote[5]{
Covariant differentiation of the general equation~(\ref{EOF1}) gives
\begin{equation}
T_{\mu\nu}^{\,\,\,\,\,;\nu}=\phi^{,\nu}L''(\phi)\Bigl(\frac{T_{\mu\nu}}{L'(\phi)}+\frac{[2L(\phi)-\phi L'(\phi)]g_{\mu\nu}}{2\kappa[L'(\phi)]^2}\Bigr),
\label{conserv}
\end{equation}
which for an isotropic universe yields~(\ref{cons}).}
In the Jordan frame such a tensor is always conserved~\cite{Niko,Koi},
although~\cite{Pal6} arrives at the different conclusion.

The generalized energy density in the late
FLRW universe with the metric~(\ref{FLRW}) is given by
\begin{equation}
\Im=\frac{3\epsilon}{4}+\frac{3\alpha}{16\kappa}+\frac{3\kappa \epsilon^2}{16\alpha},
\end{equation}
and the generalized pressure is
\begin{equation}
\wp=-\frac{3\alpha}{16\kappa}+\frac{\kappa \epsilon^2}{16\alpha}.
\end{equation}
From the conservation law~(\ref{cons})~\cite{Niko}
we obtain the scaling law for the energy density:
\begin{equation}
\epsilon a^3=E_{\infty} e^{-\frac{\kappa\epsilon}{6\alpha}},
\label{scalen2}
\end{equation}
where $E_{\infty}$ is a positive constant.
In the limit of very small $\epsilon$, the above relation simplifies to the 
usual expression for nonrelativistic matter, $\epsilon a^3=E_{\infty}$,
where $E_{\infty}$ has the meaning of a quantity proportional to 
the total energy of matter within a sphere of radius $a$.

Similarly, we obtain the energy density dependence on time:
\begin{equation}
ln\frac{\epsilon}{\epsilon_0}-\frac{11\kappa(\epsilon-\epsilon_0)}{6\alpha}=-\frac{3ct\sqrt{\alpha}}{4},
\end{equation}
where time is measured from the instant at which $\epsilon$ is equal to 
a given value $\epsilon_0$.
After some mathematical manipulations, we arrive at the following expression 
for the scaling factor as a function of time in the quadratic approximation:
\begin{equation} 
a^3(t)=\frac{E_{\infty}}{\epsilon_0}\bigl(e^{3H_{ca}t}-\frac{2\kappa\epsilon_0}{\alpha}+\frac{27\kappa^2\epsilon_0^2}{8\alpha^2}e^{-3H_{ca}t}\bigr),
\end{equation}
where $H_{ca}=c\sqrt{\alpha/16}$~\cite{Niko}.
The late universe approaches a de Sitter spacetime 
exponentially fast, as in the Jordan frame~\cite{Pal2}. 

We note that the energy density of matter scales like $a^{-3}$ 
(nonrelativistic matter) in both considered limits,
$\kappa\epsilon\ll\alpha$ and $\kappa\epsilon\gg\alpha$.
Therefore, there exists the largest deviation from such a scaling in the
present epoch when $\kappa\epsilon$ and $\alpha$ are on the same order.

\section{The present universe in $R+1/R$ gravity}
From the $00$ component of equation~(\ref{EOF1}) we can obtain the Hubble
parameter for a flat FLRW universe filled with dust, as a function of 
$\phi$~\cite{Niko}:
\begin{equation}
H(\phi)=\frac{c}{L'(\phi)}\sqrt{\frac{\phi L'(\phi)-3L(\phi)}{6}}.
\label{Hub1}
\end{equation}
For the Lagrangian~(\ref{acc1}), we find
\begin{equation}
H(\phi)=\frac{c\sqrt{-\phi}}{1+\frac{\alpha^2}{3\phi^2}}\sqrt{\frac{1}{3}-\frac{2\alpha^2}{9\phi^2}}.
\label{Hub2}
\end{equation}
Similarly, we derive the deceleration parameter $q=-\frac{a\ddot{a}}{\dot{a}^2}$ using the $11$ component:
\begin{equation}
q(\phi)=\frac{2\phi L'(\phi)-3L(\phi)}{\phi L'(\phi)-3L(\phi)}.
\label{dec1}
\end{equation}
For the Lagrangian~(\ref{acc1}), we obtain
\begin{equation}
q(\phi)=\frac{5\alpha^2-3\phi^2}{4\alpha^2-6\phi^2}.
\label{dec2}
\end{equation}
The time dependence of $q$ is determined by combining equation~(\ref{dec2})
with the function $\phi=\phi(t)$ which is an integral of~\cite{Niko}:
\begin{equation}
\dot{\phi}=\frac{c(-\phi+\frac{\alpha^2}{\phi})\sqrt{\frac{2\alpha^2}{\phi}-3\phi}}{1-\frac{2\alpha^4}{3\phi^4}+\frac{7\alpha^2}{3\phi^2}}.
\label{phidotca}
\end{equation}

We also find the expression for the apparent matter density parameter
$\Omega=\frac{\epsilon_0}{\epsilon_c}$:
\begin{equation}
\Omega_M=2L'(\phi_0)\frac{\phi_0 L'(\phi_0)-2L(\phi_0)}{\phi_0 L'(\phi_0)-3L(\phi_0)},
\label{omega1}
\end{equation}
where $\epsilon_c=\frac{3H_0^2}{\kappa c^2}$ is the critical energy density
and the subscript $0$ refers to the present time.
For the Lagrangian~(\ref{acc1}), this parameter becomes
\begin{equation}
\Omega_M=\Bigl(1+\frac{\alpha^2}{3\phi_0^2}\Bigr)\frac{3\alpha^2-3\phi_0^2}{2\alpha^2-3\phi_0^2}.
\label{omega2}
\end{equation}
Substituting $\phi_0=\phi_0(q_0)$ from equation~(\ref{dec2}) to~(\ref{omega2})
gives
\begin{equation}
\Omega_M=\frac{q_0^2-1}{q_0-\frac{5}{4}}
\label{rel1}
\end{equation}
or
\begin{equation}
q_0=\frac{1}{2}\Bigl(\Omega_M-\sqrt{\Omega_M^2-5\Omega_M+4}\Bigr).
\label{rel2}
\end{equation}

For the early universe $|\phi|\gg\alpha$ and $q\approx\frac{1}{2}$.
For the late universe $\phi\rightarrow-\alpha$ and $q\rightarrow-1$.
The transition from the decelerated ($q>0$) to accelerated ($q<0$) phase
occurs at $q=0$, which takes place when $\phi$ equals
\begin{equation}
\phi_t=-\sqrt{\frac{5}{3}}\alpha.
\label{dec3}
\end{equation}
The Hubble parameter and the energy density of matter
at the time of the deceleration-to-acceleration transition are given,
respectively, by
\begin{eqnarray}
& & H_t= c\frac{\sqrt{5}}{6}\sqrt{\sqrt{\frac{5}{3}}\alpha}, \\
& & \epsilon_t=\sqrt{\frac{5}{27}}\frac{\alpha}{\kappa}.
\label{Hub3}
\end{eqnarray}
We see that the energy density at this time is on the order of 
$\frac{\alpha}{\kappa}$, i.e. it scales with the largest deviation 
from nonrelativistic dust.

In the Jordan frame, the energy density scales like $a^{-3}$ which is
equivalent to $(1+z)^3$, where $z$ is the redshift.
The energy density in the Einstein frame is related to that in the Jordan
frame by the conformal factor $L'(\phi)$.
Therefore,
\begin{equation}
\epsilon=\epsilon_0 (1+z)^3\frac{L'(\phi)}{L'(\phi_0)}.
\label{scaling1}
\end{equation}
Substituting here $\epsilon(\phi)$ from equation~(\ref{struc3}) gives
\begin{equation}
\frac{\phi L'(\phi)-2L(\phi)}{\phi_0L'(\phi_0)-2L(\phi_0)}\,\frac{[L'(\phi_0)]^2}{[L'(\phi)]^2}=(1+z)^3,
\label{scaling2}
\end{equation}
from which we obtain $\phi=\phi(z,\phi_0)$.
In the case of the Lagrangian~(\ref{acc1}), $\phi$ is a solution
of a quintic equation.
From the redshift dependence of $\phi$ we can derive the relations $H=H(z)$
and $q=q(z)$ using equations~(\ref{Hub2}) and~(\ref{dec2}), respectively.

The observed value of the present matter density parameter is
$\Omega_M=0.29\pm^{0.05}_{0.03}$~\cite{Gold}.
Substituting this value into equation~(\ref{rel2}) gives the prediction
for the apparent deceleration parameter:
\begin{equation}
q_0=-0.67\pm^{0.06}_{0.03},
\end{equation}
which agrees with the observed $q_0=-0.74\pm0.18$~\cite{Gold}.\footnote[6]{
The predictions for non-dimensional cosmological parameters in the $R+1/R$
gravity are independent of the value of $\alpha$.}
The deceleration-to-acceleration transition happened in the past and the
universe expands with a positive acceleration.
The corresponding value of $\phi_0$ is determined by equation~(\ref{dec2}):
\begin{equation}
\phi_0=(-1.05\pm0.01)\alpha.
\label{phi0}
\end{equation}
From equation~(\ref{Hub2}), we find
\begin{equation}
\alpha=(12.5\pm^{0.4}_{0.7})\frac{H_0^2}{c^2}.
\end{equation}
The observed value $H_0=71\pm4\frac{km}{s\cdot Mpc}$~\cite{WMAP} leads to
\begin{equation}
\alpha=(7.35\pm^{1.12}_{1.17})\times10^{-52}m^{-2}.
\end{equation}

The deceleration-to-acceleration redshift $z_t$ can be obtained from
equations~(\ref{struc3}), (\ref{dec3}), (\ref{Hub3}) and (\ref{scaling1}).
Substituting the numerical value for $\phi_0$ from~(\ref{phi0}) yields
\begin{equation}
z_t=0.86\pm^{0.06}_{0.10}.
\end{equation}
This value overlaps with the $\Lambda CDM$ value of
$z_t=0.71\pm^{0.20}_{0.19}$\footnote[7]{For a flat universe in the
$\Lambda CDM$ model, $z_t$ is a simple function of $\Omega_M$:
$z_t=\Bigl[\frac{2(1-\Omega_M)}{\Omega_M}\Bigr]^{1/3}-1$.} but disagrees
with the observed $z_t=0.46\pm0.13$~\cite{Gold}.
The extraction of $z_t$ from the data appears, however, to be much less
robust than the extraction of $q_0$~\cite{Dec}.
Therefore, we need stronger measurements of $z_t$ to verify if the
$R+1/R$ gravity is a viable explanation of current cosmic acceleration.

The last non-dimensional parameter in our study is $q_1$, the coefficient
in the Taylor expansion of $q(z)$ around $z=0$: $q(z)=q_0+q_1 z$.
We use $q_1=\frac{\partial q}{\partial z}|_{z=0}=\frac{\partial q/\partial\phi|_{\phi=\phi_0}}{\partial z/\partial\phi|_{\phi=\phi_0}}$ to obtain
\begin{equation}
q_1=9\frac{L'(\phi L'^2-\phi LL''-LL')(\phi L'-2L)}{(\phi L'-3L)^2(4LL''-L'^2-\phi L'L'')}\Big|_{\phi=\phi_0}
\label{q1}
\end{equation}
from equations~(\ref{dec1}) and~(\ref{scaling2}).
The Lagrangian~(\ref{acc1}) and the current value of $\phi$~(\ref{phi0}) give
\begin{equation}
q_1=0.81\pm^{0.17}_{0.16},
\end{equation}
which slightly overlaps with the observed $q_1=1.59\pm0.63$~\cite{Gold}.

\section{Summary}
Current cosmic acceleration can be explained by adding a $1/R$ term to the 
Einstein-Hilbert Lagrangian.
We used the metric-affine variational formalism and chose the Einstein 
conformal frame to be physical.
In this frame we derived the field equations
for the early and late matter-dominated FLRW universe,
in the quadratic approximation with respect to small quantities.
We showed that the largest deviation of the energy density $\epsilon$ scaling
from the usual scaling of nonrelativistic dust occurs in the present epoch.
We did not give the exact form of the field equations for any value of
$\epsilon$, although it can be done analytically for the $R+1/R$ Lagrangian. 
Instead, we derived the expressions for non-dimensional cosmological
parameters to verify if the $R+1/R$ gravity is compatible with
observations.

We found that the current deceleration parameter $q_0$ derived from
the apparent matter density parameter $\Omega_M$ is consistent with
observations.
There is also a tiny overlap between the predicted and observed values
for the redshift derivative $q_1$ of the deceleration parameter.
The predicted redshift of the deceleration-to-acceleration transition $z_t$
agrees with that in the $\Lambda CDM$ model but it is larger than the value
estimated from SNIa observations.
Since the robustness of the $z_t$ measurements is weaker than that of $q_0$,
the question on the viability of the $R+1/R$ gravity in the Einstein frame
remains open.

\end{document}